# PKN PROBLEM FOR NON-NEWTONIAN FLUID

Alexander M. Linkov

*Institute for Problems of Mechanical Engineering, Russian Academy of Sciences,*

*61 Bol'shoy Pr. V.O., 199 178, Russia*

e-mail: voknilal@hotmail.com

***Abstract.*** The paper presents analytical solution for hydraulic fracture driven by a non-Newtonian fluid and propagating under plane strain conditions in cross sections parallel to the fracture front. Conclusions are drawn on the influence of the fluid properties on the fracture propagation.

## 1. INTRODUCTION

Hydraulic fracturing is widely used for increasing production of oil or gas wells. Because of its practical significance, it has been studied in many papers starting from that by Khristianovich and Zheltov [1,2]. The model of these authors, considered also by Geertsma and de Klerk [3] and called the KGD-model, presumes plane-strain conditions in the cross-sections orthogonal to the fracture front. It refers to the initial stage of the fracture propagation, when the influence of the fracture toughness may be essential. In contrast, the Perkins-Kern model [4], augmented by Nordgren [5] and called the PKN-model, presumes plane-strain conditions in the cross-sections parallel to the front. It refers to far longer fracture, when the major resistance to the crack propagation is caused by fluid viscosity. Below we shall focus on this stage of the fracture propagation. Most of theoretical work has focused on studying asymptotic behaviour of solutions and distinguishing parameters defining various regimes of the fracture propagation (see, e.g. reviews in papers [5-10]). Only a few papers have contained complete solutions of model problems for a finite fracture [5, 8, 11-14]. The solutions have been obtained numerically by using the net pressure and fracture opening as unknowns in quite involved calculations. Recently [15,16], it has been disclosed that the hydraulic fracture problem, when considered as a boundary value problem under fixed position of the front and zero lag, is ill-posed and needs regularization. This finding resulted in the modified formulation of the problem [17-19], which provides notable computational and analytical advantages. Specifically, the problems by Nordgren [5] and Spence and Sharp [11] could be solved analytically for a Newtonian fluid [17]. It also opens the opportunity to extend results to non-Newtonian fluids. The present paper employs this option. For certainty and having in mind vast practical applications (e. g. [20, 21]), we consider the PKN model. Our purpose is to obtain and analyze solutions, which become available, and to clearly reveal features of the fracture propagation for fracturing fluids with arbitrary behavior and consistency indices.

## 2. PROBLEM FORMULATION IN TERMS OF PARTICLE VELOCITY

We consider a viscous fluid with the power-type reological law, connecting the shear stress $\sigma_\tau$ with the shear strain rate $\dot\gamma$:

$$\sigma_\tau = M\dot\gamma^n. \qquad (1)$$

Herein, *M* is the consistency index, *n* is the exponent, called fluid behavior index. Commonly, shear-thinning fluids, for which $0 \le n < 1$, are used in practice of hydraulic fracturing. The case *n* = 1 corresponds to a Newtonian fluid with dynamic viscosity $M = \mu$; the case $n = 0$ corresponds to a perfectly plastic fluid with constant shear strength $M = \sigma_{\tau 0}$.

For a flow in a narrow channel of the width $w$, the fluid may be assumed incompressible and conventional derivation, employing (1), yields the dependence between the particle velocity $v$, averaged across the channel width $w$, and gradient of pressure:

$$v = \left(-k_f w^{n+1} \frac{\partial p}{\partial x}\right)^{1/n}, \tag{2}$$

where the coefficient $k_f$ is inversely proportional to the consistency index: $k_f = 1/(\theta M)$. For an elliptical channel with axes $w$ and $h$, the Lamb-type equation defines the factor $\theta$ (e. g. [10]): $\theta = 2\left[\frac{\pi(1+\pi n - n)}{2n}\right]^n$;

for a thin plane channel, the Poiseuille value is often used (e. g. [12]): $\theta = 2\left[\frac{2(2n+1)}{n}\right]^n$. The ratio of these factors does not differ significantly from the unit, being $12/\pi^2 \approx 1.216$ for a Newtonian fluid and 1.0 for a perfectly plastic fluid. Below, for certainty, we shall use the Poiseuille value. For it, $\theta = 12$, when the fluid is Newtonian; and $\theta = 2$, when the fluid is perfectly plastic.

By definition, the flux through the channel width is

$$q = wv, \tag{3}$$

A fracture of the height $h$ propagates in the $x$-direction (Fig. 1) in elastic rock with the elasticity modulus $E$ and the Poisson's ratio $\nu$. In accordance with the PKN-model, we assume that the crack length $x_*(t)$ is large enough to have plane strain conditions in the cross-section parallel to the fracture front. Then the dependence of the net-pressure $p$ and the opening $w$ averaged over the height is (see, e.g., [5]):

$$p = k_r w, \tag{4}$$

where $k_r = (2/\pi h)E/(1-\nu^2)$. Substitution of (4) into (2) yields for the PKN model:

$$v = \left(-\frac{k_f k_r}{n+2} \frac{\partial w^{n+2}}{\partial x}\right)^{1/n}. \tag{5}$$

At the points of the fracture front, the particle velocity equals to the front propagation speed $v_*$. Thus we have the speed equation (SE) [15]:

$$v_* = \frac{dx_*}{dt} = v(x_*). \tag{6}$$

From (5) and (6) it follows that to have the propagation with non-zero finite speed, the function $y = w^{n+2}$ should be *linear* in $x$ near the front. This suggests using the function $y$, which we call the modified opening, and the particle velocity $v$ as proper variables, instead of the opening $w$ and the net pressure $p$ [15,17]. Denoting $\alpha = 1/(n+2)$, we have $w = y^\alpha$.

In terms of the variables $y$ and $v$, the lubrication partial differential (PDF) equation reads:

$$\frac{\partial v}{\partial x} + \frac{\alpha}{y}\frac{\partial y}{\partial x}v + \frac{\alpha}{y}\frac{\partial y}{\partial t} + \frac{1}{y^\alpha}q_e = 0. \tag{7}$$

Herein, $q_e$ is the term accounting for leak-off into formation ($q_e \geq 0$); henceforth, it is assumed that the leak-off may be singular at the fluid front; still the product $y^{1-\alpha}q_e$ has to tend to zero when $x \to x_*$.

The dependence between $v$ and $y$ follows from (5):

$$v = \left(-k_f k_r \alpha \frac{\partial y}{\partial x}\right)^{1/n}. \tag{8}$$

The initial condition for the PDE (7) expresses that there is no opening along a perspective fracture path ($w(x,0)=0$). In terms of the modified opening, we have:

$$y(x,0) = 0. \tag{9}$$

There are two boundary conditions (BC) for the PDF (7), which is of second order in the spatial coordinate $x$. One of them is the condition of the prescribed influx $q_0(t)$ (per unit height) at the inlet $x = 0$. In view of (3), this condition reads:

$$y^\alpha v\big|_{x=0} = q_0(t). \tag{10}$$

The second is the condition of zero opening, and consequently, zero modified opening at the fracture front $x = x_*$:

$$y(x_*,t) = 0. \tag{11}$$

The SE (6), in view of (8), becomes:

$$v(x_*,t) = v_*(t) = \left(-k_f k_r \alpha \frac{\partial y}{\partial x}\right)^{1/n}\bigg|_{x=x_*}. \tag{12}$$

The problem is to solve the PDE (7), where the dependence between $v$ and $y$ is given by (8), under the initial condition (9) and the BC (10), (11). Besides, as shown in [16], under the BC (11), the PDE (7) in the limit $x \to x_*$ yields that the SE (12) is met identically. Hence, for a fixed position of the fracture front, we actually have two BC (11), (12) at the front rather than one condition (11). This makes the boundary value (BV) problem (7)-(11) *ill-posed* for any fixed position of the front $x_*$ [15,16]. To avoid complications when solving the PKN-problem, there are various options. One of them consists in using ε-regularization [15,16], another employs including $x_*$ as an additional dynamic unknown in a dynamic system of ODE, obtained after spatial discretization [18]. Below we employ the third option, used in [15-17] to obtain bench-mark solutions for Newtonian fluid. It consists in solving the initial value problem (7), (8), (11), (12) for a fixed front position $x_*$ and finding that speed $v_*$, for which the condition (10) at the inlet is met.

## 3. NORMALIZED VARIABLES. SELF-SIMILAR FORMULATION

We use typical values of the influx per unit height $q_n$ and time $t_n$ to normalize physical variables. The normalizing length $x_n$, opening $w_n$, modified opening $y_n$, pressure $p_n$, velocity $v_n$ and leak-off $q_{\ln}$ are defined as

$$x_n = (k_r k_f q_n^{n+2})^{\frac{1}{2n+3}} t_n^{\frac{2n+2}{2n+3}}, \quad w_n = q_n t_n / x_n, \quad y_n = w_n^{1/\alpha}, \quad p_n = k_r w_n, \quad v_n = x_n / t_n, \quad q_{\ln} = q_n / x_n. \tag{13}$$

The dimensionless variables are:

$$x_d = x/x_n, \; x_{*d} = x_*/x_n, \; t_d = t/t_n, \; v_d = v/v_n, \; v_{*d} = v_*/v_n, \; w_d = w/w_n, \; y_d = y/y_n, \tag{14}$$

$$p_d = p/p_n, \; q_d = q/q_n, \; q_{0d} = q_0/q_n, \; q_{ld} = q_l/q_{\ln}.$$

All the equations of the previous section keep their form for dimensionless variables with the change of the coefficients $k_r$ and $k_f$ in (8) and (12) to the unity. This excludes the consistency index from equations in normalized variables. When there may be no confusion, we shall assume $k_r = 1$, $k_f = 1$ and omit the subscript '$d$' in the notation of dimensionless variables.

Consider the case when the dimensionless influx at the inlet is prescribed by the power dependence on the dimensionless time:

$$q_0(t) = t^{\beta_q}, \tag{15}$$

where $\beta_q$ is a dimensionless constant. Note that $\beta_q = 0$ for constant influx.

For the dependence (15) and zero leak-off, the solution of the problem (7)-(12) may be found in terms of the self-similar variables defined by equations:

$$x = \xi t^{\beta_*}, \quad x_* = \xi_* t^{\beta_*}, \quad v = V(\xi)t^{\beta_*-1}, \quad v_* = V_* t^{\beta_*-1}, \tag{16}$$

$$w = W(\xi)t^{\beta_w}, \quad y = Y(\xi)t^{\beta_w/\alpha}, \quad p = P(\xi)t^{\beta_p}, \quad q = Y(\xi)^\alpha V(\xi) t^{\beta_q},$$

where $\xi_*$ and $V_* = \xi_* \beta_*$ are constants, expressing the self-similar fracture length and speed of propagation, respectively. As $\xi/\xi_* = x/x_*$, the self-similar coordinate $\xi = \xi_* x/x_*$ proportional to the distance normalized by the fracture length $x_*$. Thus, in fact, the formulae (16) represent the solution in the form of products of functions with separated variables $\varsigma = \xi/\xi_* = x/x_*$ and $t$. To qualitatively account for leak-off, we assume that the leak-off term is also represented in the form with separated variables $q_l = Q_l(\xi)t^{\beta_l}$. The function $Q_l(\xi)$ may be singular at the fracture front $\xi_*$, although the singularity should not be too strong: $Q_l(\xi) = o((\xi_* - \xi)^{\alpha-1})$. Substitution of (16) into (7), (10) and (12) yields that the powers of time cancel when

$$\beta_w = \beta_p = \frac{1+(n+1)\beta_q}{2n+3}, \quad \beta_* = \frac{2(n+1)+(n+2)\beta_q}{2n+3}, \quad \beta_l = \beta_w - 1. \tag{17}$$

With the choice (17), the PDE (7) becomes the ordinary differential equation (ODE) in the self-similar variables:

$$\frac{dV}{d\xi} + \alpha \frac{V(\xi) - V_* \xi/\xi_*}{Y(\xi)} \frac{dY}{d\xi} + \beta_w + \frac{1}{Y(\xi)^\alpha} Q_e(\xi) = 0. \tag{18}$$

The dependence (8), the BC (10), (11) and the SE (12) become, respectively:

$$V(\xi) = \left(-\alpha \frac{dY}{d\xi}\right)^{1/n}, \tag{19}$$

$$Y^\alpha V(0) = A, \tag{20}$$

$$Y(\xi_*) = 0, \tag{21}$$

$$V_* = \xi_* \beta_* = \left(-\alpha \frac{dY}{d\xi}\right)^{1/n}\bigg|_{\xi=\xi_*}. \tag{22}$$

Actually, in (20), $A = 1$. We have written $A$ for further discussion of the solution. The initial condition (9) for the PDE (7) is met by the representation (16) for $w$, when $\beta_w > 0$. In view of the first of (17), we have $\beta_w > 0$ when $\beta_q > -1/(n+1)$. Thus $\beta_q$ may be negative; this means that the solution may include the case of the influx decreasing in time from initially infinite value.

For any fixed $\xi_*$, the problem of solving the ODE (18), where the dependence between $V$ and $Y$ is given by (19), under the boundary conditions (20), (21) is ill-posed. Indeed, in the limit $\xi \to \xi_*$, a solution of (18), satisfying the BC (21), identically satisfies the SE (22), as well. Hence, at the point $\xi = \xi_*$ we actually have *two* rather than one conditions. Therefore, for any fixed $\xi_*$, these two *initial* conditions (21) and (22) at $\xi = \xi_*$ completely define the solution $Y(\xi)$ of the ODE (18). Therefore, they define the derivative $dY/d\xi$, as well, and consequently, $V(\xi)$ and the value $A$ in the condition (20). Thus, in accordance with the results of the papers [15,16], it is impossible to solve the *boundary value* problem (18)-(21) without regularization. Therefore, it is reasonable to solve the *initial value* (Cauchy) problem (18), (19), (21), (22) for a fixed $\xi_*$. Substitution of the solution into the BC at the inlet (20) gives the corresponding influx $A$. By changing $\xi_*$, we may find that value of $\xi_*$, for which the BC (20) is met to a prescribed tolerance when $A = 1$.

Actually, in the case of zero leak-off, there is no need in solving the problem for various $\xi_*$. It can be shown that if $Y_1(\xi)$ is the solution for $\xi_* = \xi_{*1}$ so that the corresponding influx is $A = A_1$, then the solution for an arbitrary influx $A$ is given by equation:

$$\xi_* = \xi_{*1}\left(\frac{A}{A_1}\right)^{(n+2)/(n+3)}, \quad Y(\xi) = \left(\frac{\xi_*}{\xi_{*1}}\right)^{n+1} Y\left(\xi\frac{\xi_{*1}}{\xi_*}\right),$$

Hence, it is sufficient to find the solution for $\xi_{*1} = 1$. Similar conclusion was obtained in the papers [15,16] for the particular case of Newtonian fluid ($n = 1$).

## 4. ANALYTICAL SOLUTION OF THE PROBLEM

*Generic case.* In view of the BC (21) and the SE (22), the function $Y(\xi)$ is at least linear near the fluid front $\xi_*$. Assume that the leak-off term $Q_l(\xi)$ is of order $O((\xi_* - \xi)^\alpha)$ near the front. Then we may represent $Y(\xi)$, $V(\xi)$ and $Q_l(\xi)$ by using power series in $\tau = 1 - \xi/\xi_*$:

$$Y(\xi) = \frac{\xi_*^{n+1}\beta_*^n}{\alpha}\sum_{j=1}^{\infty} a_j\tau^j, \quad V(\xi) = V_*\sum_{j=0}^{\infty} b_j\tau^j, \quad Q_l(\xi) = \tau^\alpha \sum_{j=0}^{\infty} q_j\tau^j \tag{23}$$

where the coefficients $q_j$ of leak-off are prescribed. The BC (21) and SE (22) give $a_1 = b_0 = 1$. Then the expansions (23) correspond to solving well-posed *initial value* problem (18), (19), (21), (22).

The dependence (19) yields $\sum_{k=0}^{\infty}(k+1)a_{k+1}\tau^k = \left(\sum_{j=0}^{\infty} b_j\tau^j\right)^n$, that allows us to recurrently express the coefficients $a_{k+1}$ ($k = 1,\ldots$) via $b_j$ ($j = 1,\ldots,k$). For the first five coefficients we have:

$$a_1 = b_0 = 1, \quad a_2 = \frac{1}{2}nb_1, \quad a_3 = \frac{1}{6}n[(n-1)b_1^2 + 2b_2], \quad a_4 = \frac{1}{24}n[(n-1)(n-2)b_1^3 + 6(n-1)b_1b_2 + 6b_3],$$

$$a_5 = \frac{1}{120}n[(n-1)(n-2)(n-3)b_1^4 + 12(n-1)(n-2)b_1^2b_2 + 24(n-1)b_1b_3 + 24b_4]. \tag{24}$$

The coefficients $a_k$ decrease faster than $1/k^2$ with growing $k$. Substitution of the series (23) into the ODE (18) gives the second group of recurrent equations for $j \geq 2$:

$$b_j = -\frac{1}{j+\alpha}\left\{\sum_{k=2}^{j}(j-k+1+\alpha k)a_kb_{j-k+1} + (\alpha j - \frac{\beta_w}{\beta_*})a_j - C_l\sum_{k=1}^{j} c_k q_{j-k}\right\}, \tag{25}$$

with the starting values $a_1 = b_0 = 1$, $b_1 = \frac{1}{1+\alpha}\left(-\alpha + \frac{\beta_w}{\beta_*} + C_l q_0\right)$. Herein, $C_l = \left(\frac{\alpha}{\xi_*^{n+1}\beta_*^{n+1/\alpha}}\right)^\alpha$ and the coefficients $c_k$ are recurrently evaluated via $a_i$ ($i = 1,\ldots,k$) from equation $\sum_{k=1}^{\infty} c_k\tau^k = \tau\left(\sum_{i=0}^{\infty} a_{i+1}\tau^i\right)^\alpha$. For the first five coefficients it gives equations similar to (24):

$$c_1 = a_1 = 1, \quad c_2 = (1-\alpha)a_2, \quad c_3 = \frac{1}{2}(1-\alpha)[-\alpha a_2^2 + 2a_3], \quad c_4 = \frac{1}{6}(1-\alpha)[\alpha(\alpha+1)a_2^3 - 6\alpha a_2a_3 + 6a_4],$$

$$c_5 = \frac{1}{24}(1-\alpha)[-\alpha(\alpha+1)(\alpha+2)a_2^4 + 12\alpha(\alpha+1)a_2^2 a_3 - 24\alpha a_2 a_4 + 24a_5]. \tag{26}$$

Starting from $a_1 = b_0 = c_1 = 1$, we find $b_1$; then $a_2$ is found from the second of (24) and $c_2$ from the second of (26). Then (25) provides $b_2$, the third of (24) gives $a_3$, the third of (26) gives $c_3$, and so on. In the case of a Newtonian fluid ($n = 1$, $\alpha = 1/3$), we have $a_j = b_{j-1}/j$ ($j = 1,...$), and for a constant influx ($\beta_q = 0, \beta_w = 1/5, \beta_* = 4/5$) and zero leak-off ($q_{k-1} = c_k = 0$, $k = 1,...$), the recurrence formulae (25) reduce to those derived in [17].

For perfectly plastic fluid ($n = 0$, $\alpha = 1/2$), all the coefficients $a_k$, $c_k$ are zero for $k > 1$. Then the solution for a constant influx ($\beta_q = 0$) is:

$$\xi_* = (9/8)^{1/3},\ Y(\xi) = 2(\xi_* - \xi),\ V(\xi) = V_*\left[1 + \frac{1}{\beta_*\sqrt{2\xi_*}}\sum_{j=1}^{\infty}\frac{2}{2j+1}q_{j-1}\left(1 - \frac{\xi}{\xi_*}\right)^j\right],\ V_* = \frac{2}{3}\xi_*. \tag{27}$$

From (27) we see that the function $Y(\xi)$ is linear in the self-similar distance from the inlet; for zero leak-off, the self-similar velocity is constant in the entire fracture being equal to the fracture speed.

*Perfectly plastic fluid.* Above we have obtained the solution (27) for a perfectly plastic fluid in series under the assumption that leak-off is not singular at the fracture tip. Meanwhile, for a perfectly plastic fluid, more general solution may be obtained without series expansions, in quadratures. In view of the importance of this particular case, we present the solution.

For $n = 0$, the equation (19) immediately yields the first of equations (27). Substitution of $Y(\xi)$ into the self-similar lubrication equation (18) makes it a linear ODE of the first order in the self-similar velocity $V(\xi)$. Its solution, satisfying the SE (22), is:

$$V(\xi) = V_*\left[1 + \frac{1}{\beta_*\sqrt{2\xi_*\tau}}\int_0^\tau Q_l(\xi(\tau))d\tau\right], \tag{28}$$

where, as above, $\tau = 1 - \xi/\xi_*$. Note that summation of the series in the third of (27) gives the same result (28) while now leak-off may be singular at the fracture front as $o(\tau^{-\delta})$ with $0 < \delta < 0.5$.

The self-similar fracture length $\xi_*$ and correspondingly the self-similar speed $V_* = \xi_*\beta_*$ are found from the BC (20) with $A = 1$. This yields the cubic equation in $\sqrt{\xi_*}$: $a(\sqrt{\xi_*})^3 + b(\sqrt{\xi_*})^2 = 1$, where $a = \beta_*/\sqrt{\alpha}$, $b = \int_0^1 Q_l(\xi(\tau))d\tau$. A solution of the cubic equation is easily found by introducing a new variable $z_1 = a^{1/3}\sqrt{\xi_*}$ or $z_2 = b^{1/2}\sqrt{\xi_*}$. Then we obtain the equation $z_1^3 + b_1 z_1^2 = 1$ linear in $b_1 = ba^{-2/3}$, when using $z_1$, or the equation $a_2 z_2^3 + z_2^2 = 1$ linear in $a_2 = ab^{-3/2}$, when using $z_2$. The first option is convenient for small or moderate leak-off ($b_1$ is less or of order of the unity); the second option is convenient for moderate or large leak-off ($a_2$ is less or of order of the unity). Noting that equality $b_1 = 1$ implies that $a_2 = 1$, we see that the ranges of applicability of these solutions overlap. Hence, any of them may serve to easily find $\xi_*$ for moderate values of $b_1$ and $a_2$.

The value $b_1 = 0$ corresponds to negligible leak-off; then $\xi_* = a^{-2/3} = \sqrt[3]{\alpha/\beta_*^2}$. For perfectly plastic fluid, $\alpha = 1/2$, $\beta_* = 2/3$ and for small leak-off we have $\xi_* = \sqrt[3]{9/8} = 1.0400$. This value is acceptable for $\int_0^1 Q_l(\xi(\tau))d\tau < 0.33$ with the relative error not exceeding 10 percent.

The value $a_2 = 0$ corresponds to dominating influence of leak-off; then $\xi_* = b^{-1} = \left(\int_0^1 Q_l(\xi(\tau))d\tau\right)^{-1}$. To the error not exceeding 10 percent, this equation is applicable for large leak-off, when $\int_0^1 Q_l(\xi(\tau))d\tau \geq 4.1$; the corresponding $\xi_*$ is less than 0.22.

## 5. DISCUSSION OF THE RESULTS

We are interested in comparing fluid flow and fracture propagation for fluids with various behavior index $n$. Actually it sufficient to consider perfectly plastic ($n = 0$) and Newtonian ($n = 1$) fluids: the results for thinning fluids ($0 < n < 1$) are intermediate between those for these limiting cases. For certainty, we neglect leak-off ($q_l = 0$) and assume that influx is constant ($\beta_w = 0$).

The most striking general feature of the flow with zero leak-off is that the particle velocity $v$ and gradient of the modified opening $y = w^{1/\alpha}$ are practically constant along the fracture at any time instant. It can be clearly seen from Fig. 2, presenting the ratio $v(x,t)/v_*(t)$, and Fig. 3, presenting the ratio $y(x,t)/y(0,t)$, for the limit cases of perfectly plastic and Newtonian fluids. (As mentioned, the results for thinning fluids are intermediate between these two). The figures evidently show advantages of using the particle velocity $v$ and the modified opening $y = w^{1/\alpha}$ rather than the net pressure $p$ and the opening $w$ itself.

Fig. 2 and the definition of $y$ imply that the approximate solutions for the opening and pressure are

$$\frac{w(x,t)}{w(0,t)} = \frac{p(x,t)}{p(0,t)} = \left(1 - \frac{x}{x_*}\right)^\alpha. \tag{29}$$

The distribution of the flux $q$, defined by (3), is also similar to (29), because the particle velocity is almost constant along the fracture. By definitions (16), we have $v_*(t) = V_* t^{\beta_*-1}$, $w(0,t) = W_0 t^{\beta_w}$, $p(0,t) = P_0 t^{\beta_w}$, where the constants are defined by equations $V_* = \xi_*\beta_*$, $W_0 = P_0 = \left(C_Y \sum_{j=1}^{\infty} a_j\right)^\alpha$. These constants do not differ significantly for thinning fluids with various behavior indices. Specifically, in the case of constant influx ($\beta_q = 0$), we have for a perfectly plastic fluid ($n = 0$, $\alpha = 1/2$, $\beta_* = 2/3$, $\beta_w = 1/3$): $\xi_* = 1.04004$, $V_* = 0.69336$, $C_Y = 2.08008$, $W_0 = 1.44225$;

for a Newtonian fluid ($n = 1$, $\alpha = 1/3$, $\beta_* = 4/5$, $\beta_w = 1/5$): $\xi_* = 1.00101$; $V_* = 0.75398$, $C_Y = 2.40485$, $W_0 = 1.32628$.

These results imply that to an error not exceeding 4.5 percent, we may use the mean values $\xi_* = 1.02$, $V_* = 0.72$, $W_0 = 1.38$ for any thinning fluid. To this accuracy, in terms of the dimensionless variables,

8normalized in accordance with (13), (14), we obtain the simple universal analytical solution for an arbitrary thinning fluid:

$$x_{*d}(t_d) \approx 1.02 t_d^{\beta_*}, \quad v_d(x_d,t_d) \approx v_{*d}(t_d) \approx 1.02\beta_* t_d^{\beta_*-1}, \quad w_d(x_d,t_d) = p(x_d,t_d) \approx 1.38\left(1-\frac{x}{x_*}\right)^{\alpha} t_d^{\beta_w}. \quad (30)$$

From (30) we see that in the normalized variables the fracture length, particle velocity, speed of propagation, opening and pressure behave quite similarly. The difference is actually only in the exponents in time depending factors. The time exponents for a perfectly plastic fluids are $\beta_*=2/3$, $\beta_w=1/3$, $\alpha=1/2$; for a Newtonian fluid $\beta_*=4/5$, $\beta_w=1/5$, $\alpha=1/3$. Therefore, the difference in corresponding exponents for thinning fluids does not exceed 2/15 both for $\beta_*$ and $\beta_w$, it is 1/6 for $\alpha$.

For the dimensional (physical) values from (30) and the definitions (13) and (14) it follows:

$$x_*(t) = 1.02(k_f k_s q_0^{n+2})^{\beta_w} t^{\beta_*}, \quad v_*(t) = \beta_* x_*(t) t^{-1}, \quad w(x,t) \approx 1.38\left(1-\frac{x}{x_*}\right)^{\alpha}\left(\frac{q_0^{n+1}t}{k_f k_s}\right)^{\beta_w}, \quad p(x,t) = k_s w(x,t).$$

These dependences account for the influence of the consistency index entering $k_f$ and the elasticity modulus of embedding rock entering $k_s$.

**CONCLUSIONS**

The conclusions of the paper are as follows.

(i) It is confirmed that using the modified lubrication equation in proper variables provides significant analytical advantages. The variables include the particle velocity $v$ and the modified opening $y = w^{\alpha}$ that is the opening taken in the degree, which guarantees that the velocity is non-singular and non-zero at the fluid front. For the considered PKN model, the analytical solution is presented by rapidly converging series for an arbitrary behavior index. In the particular case of a perfectly plastic fluid, the solution is especially simple and expressed by quadratures.

(ii) The solution discloses important general features of hydraulic fracturing with various thinning fluids. Specifically, for zero leak-off, the particle velocity is practically constant, while the modified opening is almost linear along the fracture. The self-similar fracture length $\xi_*$ is also practically independent on the fluid behavior index $n$ ($\xi_*=0.040$ for $n = 0$; $\xi_*=1.001$ for $n = 1$). This implies that the analytical dependencies (30) of the normalized quantities on the normalized time are universal regardless of a particular thinning fluid. The differences occur mostly in the exponents of time, entering as multipliers to the normalized fracture length and opening. The difference in the exponents $\beta_*$, $\beta_w$ for various fluids is not too great: the maximal difference is 2/15 both for $\beta_*$, $\beta_w$ when comparing the limiting cases of a perfectly plastic ($n = 0$) and Newtonian ($n = 1$) fluids.

*Acknowledgement.* The author gratefully acknowledges the support of the Russian Fund of Basic Researches (Grant № 12-05-00140).

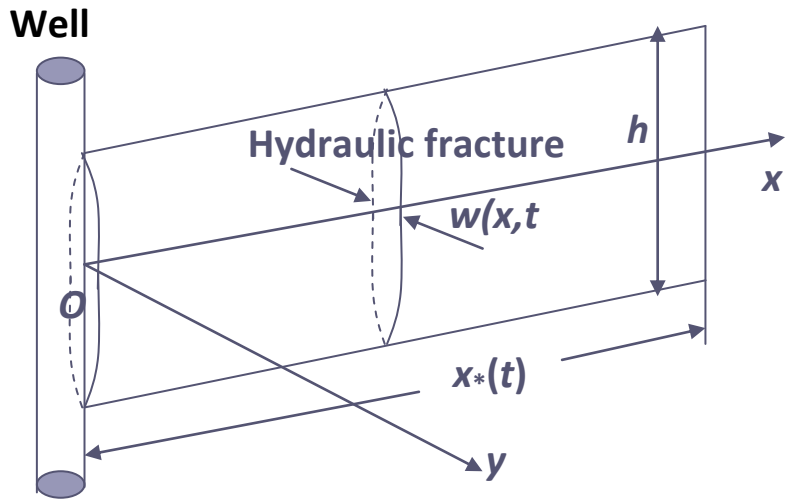

Fig. 1  Scheme of the PKN model

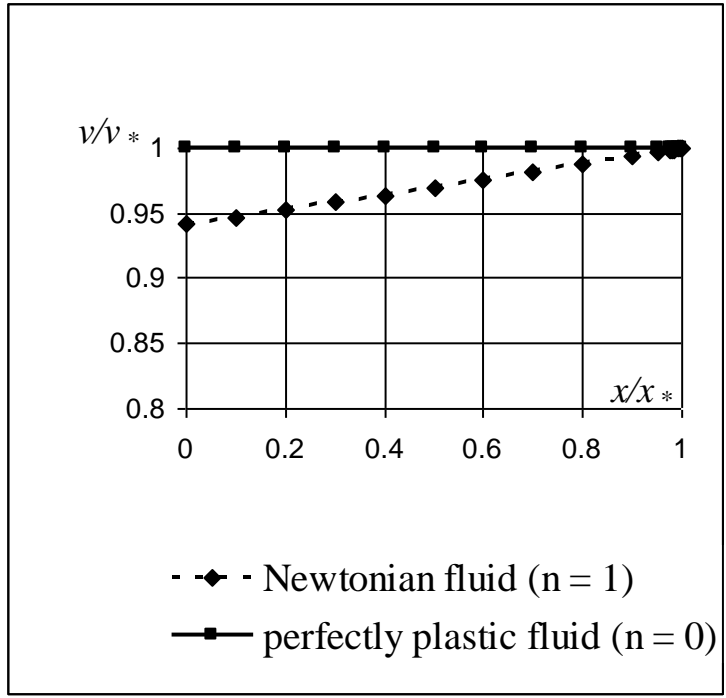

Fig. 2  Velocity distribution along the fracture

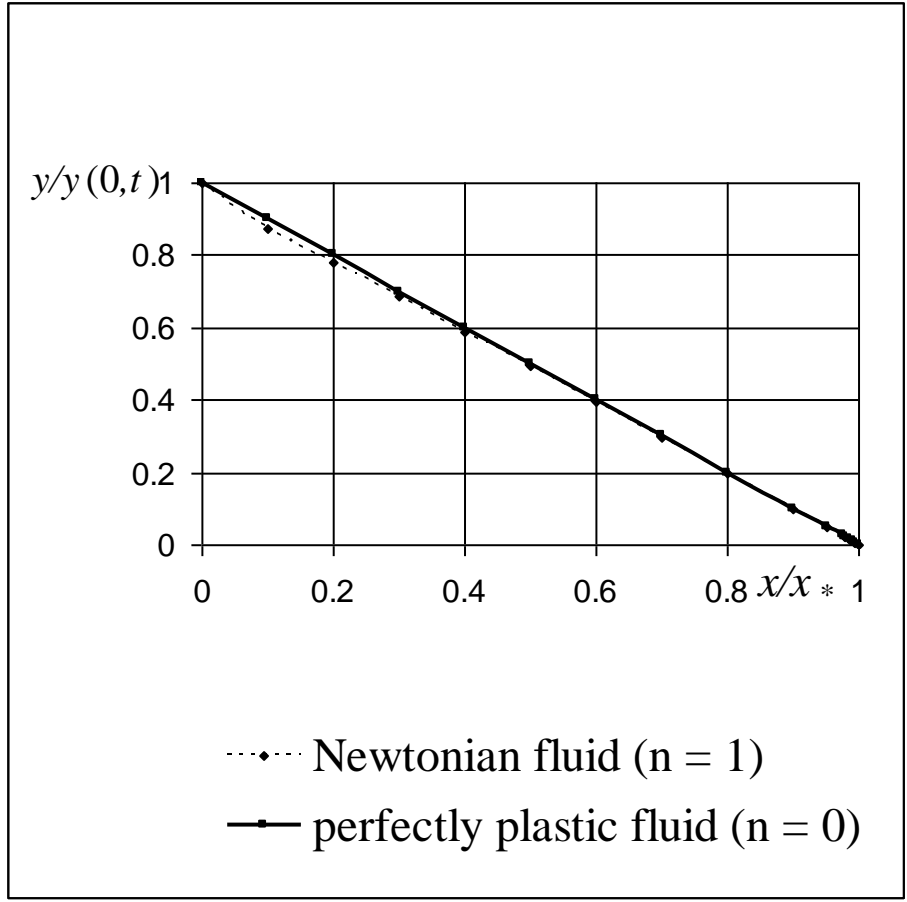

Fig. 3 Modified opening distribution along the fracture